\documentclass[aps,prl,reprint,showpacs,preprintnumbers,amsmath,amssymb,amsfonts,superscriptaddress,floatfix,showkeys]{revtex4-1}

\usepackage{graphicx}
\usepackage{dcolumn}
\usepackage{bm}
\usepackage[usenames]{color}% Include figure files

\begin{document}
\rmfamily

\title{Towards the identification of a quantum critical line \\ in the ($p$, $B$) phase diagram of CeCoIn$_5$}

\author{S. Zaum}
\affiliation{Institut f\"{u}r Festk\"{o}rperphysik, Karlsruhe Institute of Technology, D-76021 Karlsruhe, Germany}%
\affiliation{Physikalisches Institut, Karlsruhe Institute of Technology, D-76031 Karlsruhe, Germany}%
\author{K. Grube}
\affiliation{Institut f\"{u}r Festk\"{o}rperphysik, Karlsruhe Institute of Technology, D-76021 Karlsruhe, Germany}%
\author{R. Sch\"{a}fer}
\affiliation{Institut f\"{u}r Festk\"{o}rperphysik, Karlsruhe Institute of Technology, D-76021 Karlsruhe, Germany}%
%\affiliation{Physikalisches Institut, Karlsruhe Institute of Technology, D-76031 Karlsruhe, Germany}%
\author{E. D. Bauer}
\affiliation{Los Alamos National Laboratory, Los Alamos, NM 87545, USA}%
\author{J. D. Thompson}
\affiliation{Los Alamos National Laboratory, Los Alamos, NM 87545, USA}%
%\affiliation{Institut f\"{u}r Festk\"{o}rperphysik, Karlsruhe Institute of Technology, D-76021 Karlsruhe, Germany}%
\author{H. v. L\"{o}hneysen}
\affiliation{Institut f\"{u}r Festk\"{o}rperphysik, Karlsruhe Institute of Technology, D-76021 Karlsruhe, Germany}%
\affiliation{Physikalisches Institut, Karlsruhe Institute of Technology, D-76031 Karlsruhe, Germany}%
\date{\today}

\begin{abstract}
The low-temperature thermal expansion of CeCoIn$_5$ single
crystals measured parallel and perpendicular to magnetic fields
$B$ oriented along the $c$ axis yields the volume thermal-expansion 
coefficient $\beta$. Considerable deviations of $\beta(T)$
from Fermi-liquid behavior occur already within the
superconducting region of the $(B,T)$ phase diagram and become
maximal at the upper critical field $B^0_{c2}$. However,
$\beta(T)$ and the Gr\"{u}neisen parameter $\Gamma$ are
incompatible with a quantum critical point (QCP) at $B^0_{c2}$,
but allow for a QCP shielded by superconductivity and extending
to negative pressures for $B < B^0_{c2}$. Together with
literature data we construct a tentative ($p$, $B$, $T$) phase
diagram of CeCoIn$_5$ suggesting a quantum critical line in the
$(p,B)$ plane.
\end{abstract}

%\keywords{Suggested keywords}%Use showkeys class option if keyword display desired}
\maketitle

Many Ce-based heavy-fermion compounds exhibit unconventional
superconductivity on the verge of long-range magnetic order, i.e.
in the vicinity of magnetic quantum critical points
(QCPs) \cite{Monthoux_sc_review,Pleiderer_SC_f}. The presence of
these magnetic instabilities is manifested by profound deviations
from Fermi-liquid (FL) behavior \cite{HvL_all}. CeCoIn$_5$ is a
typical representative of these systems with pronounced
non-Fermi-liquid (NFL) behavior at temperatures above its
superconducting transition at $T_c = 2.3\,$K \cite{Petrovic}. With
increasing magnetic field $B$ the deviations from FL behavior
become maximal at the upper critical field
$B^0_{c2}=B_{c2}(T\rightarrow 0)\approx 5\,$T for $B\parallel c$
suggesting that NFL behavior might originate from a nearby
field-induced antiferromagnetic (AF) 
QCP at $B^0_{c2}$. Indeed, the low-temperature
dependence of the specific-heat coefficient $C/T$ at $B^0_{c2}$
follows the Hertz-Millis-Moriya (HMM) theory 
\cite{Hertz,Millis,Moriya} of a two-dimensional
spin-density-wave instability, as it might be expected for the
layered, tetragonal crystal structure of
CeCoIn$_5$ \cite{Petrovic,Bianchi_QCP}. The linear
thermal-expansion coefficient along the $c$ axis
$\alpha_c$ \cite{DonathDimSn}, the resistivity, and the thermal
conductivity \cite{Paglione}, however, reveal at $T\leq 0.3\,$K
deviations from HMM behavior that were attributed to a
dimensional crossover for $T \rightarrow 0$ \cite{DonathDimSn}.
Hall-effect measurements indicate a QCP at the second-order phase
boundary of the recently discovered Q
phase \cite{Bianchi_FFLO,Kumagai_FFLO_c}, 
possibly indicating that critical
fluctuations of the Q phase produce NFL behavior. In contrast,
resistivity measurements point to a QCP at negative hydrostatic
pressures $p$ \cite{Sidorov_p_neg} or emerging from a pressure- 
and field-dependent critical line that is masked by
superconductivity \cite{Ronning_QCP_p}.
%%%%%%%%%%%%%

% why thermal expansion and not M(T)
To search for the origin of NFL behavior of CeCoIn$_5$ in the
$(p,B)$ plane, we have exploited a characteristic QCP feature: the
accumulation of entropy at finite temperatures due to the
instability of the ground state \cite{GarstSign}. It can be
accessed from the derivatives $\partial S/\partial p$, or $\partial
S/\partial B$, the volume thermal-expansion coefficient
$\beta=-V^{-1}\cdot\partial S/\partial p$ ($V=\text{molar
volume}$) and the magnetization $\partial M/\partial T=\partial
S/\partial B$, respectively. Measurements of $M$ point to quantum
criticality directly at or below $B^0_{c2}$ \cite{Tayama_M}. 
In the vortex phase, however, $M$ cannot be used to determine 
$\partial S/\partial B$ due to irreversible flux motion. 
The thermal expansion, on the other
hand, is less sensitive to pinning effects because of the small nearly
pressure-independent entropy of the vortex lattice.

%dilatometer, sample preparation and quality, and PPMS measurments
%\section{Experimental methods}
To study $\partial S/\partial p$, we performed
thermal-expansion measurements in magnetic fields of up to
$B=14\,$T with $B\parallel c$ between $40\,$mK 
and $4\,$K.
As an important extension to previous 
measurements \cite{Oeschler,DonathDimSn} our capacitive dilatometer 
allows for thermal expansion and magnetostriction measurements 
both parallel and perpendicular to the applied magnetic field
with a resolution of $\Delta l=10^{-3}\,$\AA. 
%Sample preparation
CeCoIn$_5$ single crystals were grown in In flux. The plate-like 
crystals have typical dimensions of $l_a \approx 3\,$mm and $l_c \approx
0.5\,$mm thus enabling a relative resolution of $\Delta l/l \approx 10^{-9}$. 
To obtain the Gr\"uneisen parameter $\Gamma$ at high
fields, we measured the specific heat $C(T)$ between $350\,$mK and
$5\,$K at $B=10\,$T and $14\,$T with a Physical Properties
Measurement System from Quantum Design.

\begin{figure}[t!]
\includegraphics{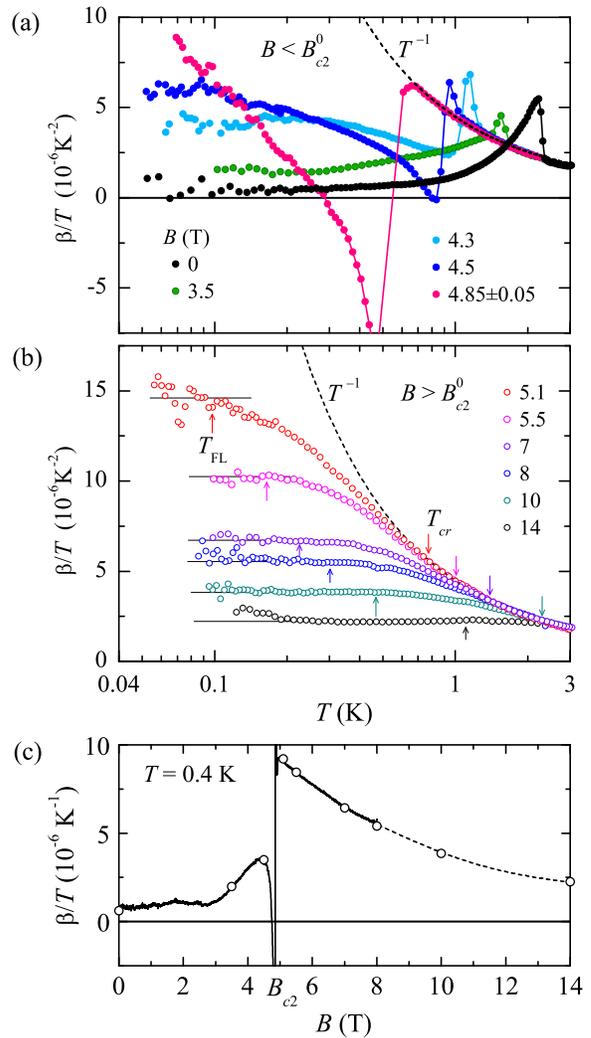}
\caption{\label{fig:alles} (Color online) The volume
thermal-expansion coefficient of CeCoIn$_5$ divided by
temperature $\beta/T$ in magnetic fields parallel to the $c$ axis
below (a) and above $B^0_{c2}$ (b). Note that the ``overshooting``
at $T_c$ is due to the fact that different samples with slightly 
different $B_{c2}(T)$ curves were used to measure $\alpha_a$ and
$\alpha_c$. The data for $B=4.85\,$T are a combination of 
measurements at $B=4.8\,$T and $4.9\,$T. The dashed lines
represent the non-Fermi-liquid behavior for the HMM model. (c)
$\beta/T$ as a function of $B$ at $T=0.4\,$K obtained from
magnetostriction measurements (continuous line) and from
temperature-dependent measurements at constant $B$ (circles).
Dashed line is a guide to the eye.}
\end{figure}

%\section{Results and discussion}
%\subsection{thermal-expansion measurements}
In a Fermi liquid the thermal-expansion coefficient approaches a
linear temperature dependence for $T\rightarrow 0$, proportional
to the specific heat. The proportionality constant is essentially
the Gr\"uneisen parameter $\Gamma=V/\kappa_s\cdot \beta/C$ 
($\kappa_s$ = adiabatic compressibility) and reflects the volume
dependence of the Fermi energy. To expose deviations from FL
behavior in CeCoIn$_5$, we plot the volume thermal-expansion coefficient, 
calculated from the measured linear thermal-expansion coefficients
$\beta=2\alpha_a + \alpha_c$, as $\beta/T$ in Fig.~\ref{fig:alles}(a) 
and (b) for fields below and above $B^0_{c2}$, respectively. 
In zero magnetic field, $\beta/T$ drops asymptotically towards zero
for decreasing $T$, just as $C/T$, due to the presence of the
superconducting gap in the energy spectrum of the quasiparticles.
At higher fields, still in the field range of the second-order
superconducting phase transition at $B_{c2}$, $\beta/T$ grows and
starts to diverge for $T\rightarrow 0$. To our knowledge, this is the
first time that such a behavior is observed within the vortex
phase of a superconductor, where, usually, the remanent Fermi
liquid in the vortex cores leads to a finite, constant $\beta/T$.
This NFL behavior corresponds to that observed in specific-heat
measurements where the entropy conservation requires a
continuously increasing $C/T\propto -\log T$ of the normal
conducting background for $T\rightarrow 0$ \cite{Petrovic,HvL_all}.

\begin{figure}[t]
\includegraphics{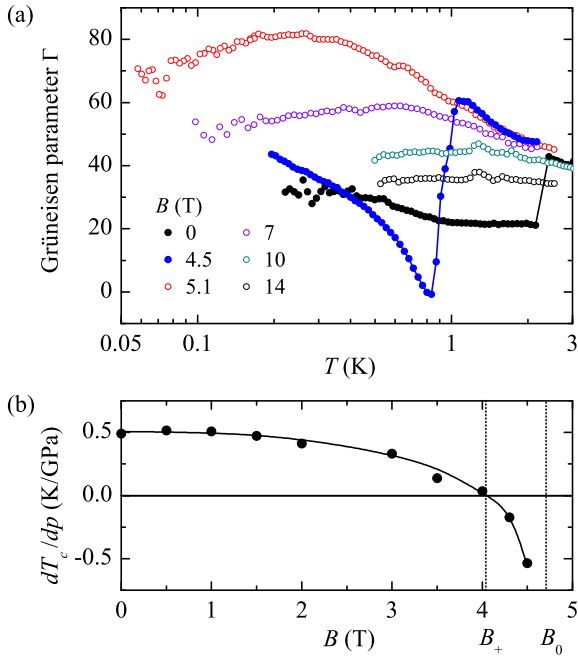}
\caption{\label{fig:GamEhr} (Color online) (a) The Gr\"uneisen
parameter $\Gamma$ versus $T$ at different constant fields
parallel to the $c$ axis. (b) The hydrostatic pressure dependences 
$dT_c/dp$ as a function of $B$. Lines are guides to the eye.}
\end{figure}

$\beta/T$ exhibits the strongest low-temperatures upturn close to
$B^0_{c2}$ [Fig.~\ref{fig:alles}(a)]. Further enhanced fields
weaken the divergence and lead, in a growing temperature range at
$T<T_\text{FL}$, to the restoration of the Fermi liquid, i.e.
$\beta/T = \text{const}$ [Fig. 1(b)]. 
This clear feature does not support an interpretation \cite{DonathDimSn} 
in terms of a dimensional crossover. 
We emphasize that at low temperatures $\beta/T$ remains positive in
the entire measured field range, except for the peaked anomaly of
the first-order transition [Fig.~\ref{fig:alles}(c)]. 
This is an important result, because the absence of a sign change 
demonstrates that $S$ has no maximum as a function of $p$ at 
$B^0_{c2}$ but grows further under negative, hydrostatic pressure 
or, equivalently, with increasing unit-cell volume. These results 
speak against a QCP in the vicinity of $B^0_{c2}(p=0)$.

%\subsection{Gr\"{u}neisen parameter}
A direct proof for the existence of a pressure-tuned QCP is the
Gr\"{u}neisen parameter $\Gamma$. If a system is dominated
by a single characteristic energy scale $E^*(V)$, $\Gamma$
reflects its volume dependence: $\Gamma=-\partial \ln E^*/ \partial
\ln V$. Since at a QCP $E^*$ vanishes, $\Gamma$ is expected to
diverge when the QCP is approached \cite{Zhu_Gruen}. We estimate
$\Gamma$ by assuming $\kappa_s$ to be roughly equal to the
isothermal compressibility of CeCoIn$_5$ $\kappa_T= 1.31\cdot
10^{-2}\,$GPa$^{-1}$ \cite{Normile} and with specific-heat data
from \cite{Bianchi_first_order,Bianchi_QCP,Ikeda} and our
measurements. The resulting $\Gamma(T)$ curves, shown in
Fig.~\ref{fig:GamEhr}(a), exhibit high $\Gamma(T)$ values 
up to 80, typical for heavy-fermion systems \cite{Visser_Gruen}, 
that illustrate their high sensitivity to volume changes. 
The $\Gamma(T)$ curve at $B=5.1\,\text{T }\approx B^0_{c2}$ is 
the highest of all fields. Here, $\Gamma$ initially 
shows a rise with decreasing $T$, but levels off below
$\approx 260\,$mK and drops again. Although the strongly enhanced
$\Gamma$ points to a pressure-tuned QCP, the lack of a
low-temperature divergence provides clear evidence against a QCP
at ambient pressure in the entire field range up to 14 T. 

To clarify a possible link between superconductivity and quantum 
criticality in CeCoIn$_5$, we compare the volume dependence of 
$T_c$ with the measured Gr\"uneisen parameter. If superconductivity 
provides the dominant energy scale $E^*\propto T_c$, $\Gamma$ is 
determined by $-d\ln T_c/d\ln V=(T_c\kappa_s)^{-1}\cdot dT_c/dp$ 
at $T << T_c$. The hydrostatic pressure dependence
$dT_c/dp=2dT_c/d\sigma_a+2dT_c/d\sigma_c$ is
calculated from the uniaxial pressure($\sigma_i$) dependences 
using the Ehrenfest relation $dT_c/d\sigma_i = T_c V\cdot \Delta
\alpha_i/\Delta C$ ($i$ = $a$, $c$). Here, $\Delta \alpha_i$ and
$\Delta C$ are the discontinuities of the linear
thermal-expansion coefficients and the specific heat at $T_c$,
respectively. The resulting $dT_c/dp$ data are displayed as a
function of $B$ in Fig.~\ref{fig:GamEhr}(b). They are in good
agreement with experiments under hydrostatic pressure
\cite{Miclea}. $dT_c/dp$ shows a sign change at $B_+\approx 4.1\pm
0.2\,$T. At higher fields, when the transition becomes
first-order, $dT_c/dp$ remains negative up to $B^0_{c2}$ as
demonstrated by the negative volume discontinuities at $T_c$ 
[Fig.~\ref{fig:alles}(a)], which is expected to produce a likewise
negative low-temperature Gr\"uneisen parameter in the field range
$B_+ < B < B^0_{c2}$. Experimentally, however $\Gamma(T)$ exhibits
large positive values in the entire field
range for $T\rightarrow0$, apart from the small field internal 
of the first-order transition near $B_{c2}$ 
[Fig.~\ref{fig:alles}(c)].
Therefore critical fluctuations of superconductivity or the
first-order transition of the Q phase cannot be responsible for
the NFL behavior.

\begin{figure}[t!]
\includegraphics{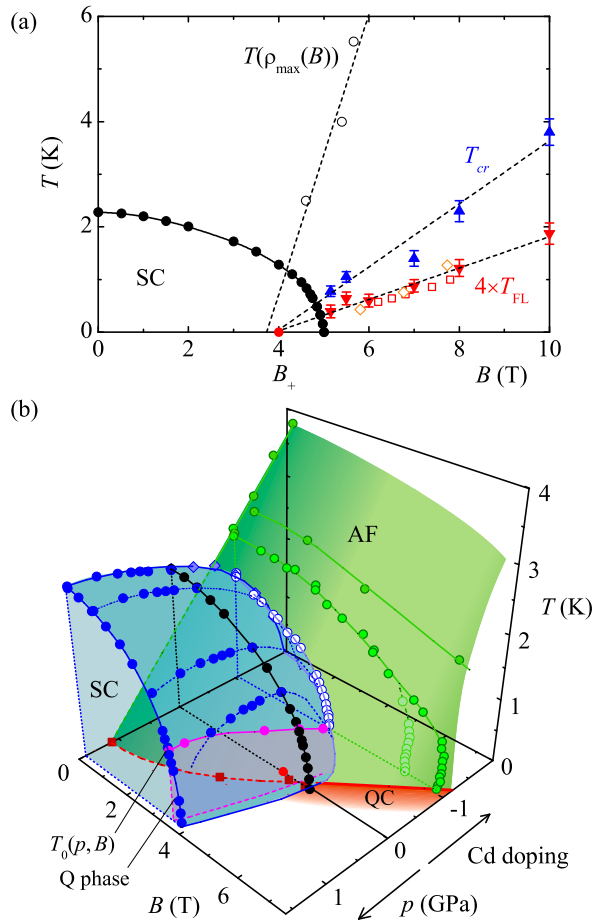}
\caption{\label{fig:PD} (Color online) (a) The ($B$, $T$) phase
diagram of CeCoIn$_5$ at ambient pressure constructed from our
measurements of the superconducting transition 
($\bullet$), the onset of FL behavior at $T_{FL}$
(\textcolor{Blue}{$\blacktriangle$}), and the change in the
critical behavior at $T_{cr}$ 
(\textcolor{Red}{$\blacktriangledown$}). In addition, literature
data are shown for the maximum of the magnetoresistance
$T(\rho_{max}(B))$ ($\circ$ \cite{Paglione_Smax}) and $T_{FL}$
determined by Hall effect (\textcolor{Red}{$\square$} \cite{Singh_TFL})
and resistivity measurements (\textcolor{Orange}{$\diamond$} 
\cite{Ronning_Hc_QCP}). (b) Tentative ($p$, $B$, $T$) phase diagram of 
CeCoIn$_5$. The data from (a)
have been extended by published measurements under
hydrostatic \cite{Tayama_To_Bc2_p,Knebel_To_Bc2_p,Lengyel_Bc2_p,Ronning_QCP_p}
and under negative, chemical pressure on
CeCoIn$_{5-x}$Cd$_x$ \cite{Donath_Cd,Tokiwa_To_Cd,Nair_Cd_B}. 
To account for the different sample qualities and varying $T_c$
definitions the measurements have been scaled to match in the
regions of overlap. $B_+$  
(\textcolor{Red}{$\bullet$}), and the onset
of the NFL behavior (\textcolor{BrickRed}{$\blacksquare$} \cite{Ronning_QCP_p}) 
are projected onto the $T=0$ plane.}
\end{figure}

% location of the QCP as a function of B)
%\subsection{Phase diagram}
The field dependence of the FL recovery [Fig.~\ref{fig:alles}(b)] 
can be used to estimate a possible field-induced origin of the 
NFL behavior in CeCoIn$_5$ at $p=0$. 
To look for such a behavior with a decrease of
$T_{FL}$ toward a QCP, we construct from our data the ($B$, $T$)
phase diagram in Fig.~\ref{fig:PD}(a). Within the experimental
error, $T_{FL}$ agrees with the onset of the $T^2$ dependence of
the resistivity $\rho$ \cite{Ronning_Hc_QCP,Paglione} 
or of constant $C/T$ values \cite{Bianchi_QCP}. Other
characteristic features that are related to the FL recovery such
as the minimum of the differential Hall
coefficient \cite{Singh_TFL}, coincide likewise with $T_{FL}$. A
striking feature of the phase diagram is that $T_{FL}$ does not
vanish at $B^0_{c2}$. For $B \leq 10\,$T, $T_{FL}$
extrapolates linearly to a $T = 0$ critical field of
$B_c=4.1\pm 0.1\,$T, which is equal to $B_+$ where, according to
$dT_c/dp=0$, $T_c$ exhibits a maximum as a function of $p$.
This suggests that the (hidden) AF QCP at $p=0$ is expected to 
occur at the field where $T_c(p)$ of the superconducting dome 
is maximal.
Furthermore, the evolution of the coherence maximum
of the magnetoresistance $\rho(B)$ with temperature, $T(\rho_{max}(B))$, 
[Fig.~\ref{fig:PD}(a)] \cite{Paglione_Smax}
extrapolates to roughly the same $B_c$ for $T\rightarrow 0$. 
Therefore a QCP at $B_c$ has been postulated \cite{Singh_TFL}. 
In the $T=0$ plane with two control parameters $p$ and 
$B$, a line of QCPs is highly plausible. This is supported by the
thermal-expansion measurements above $T_c$
[Fig.~\ref{fig:alles}(a)] or above $T_{FL}$
[Fig.~\ref{fig:alles}(b)] where all $\beta/T$ curves merge into
the same critical $T^{-1}$ dependence. Therefore, the NFL
behavior at different fields is likely to have a common origin.
The NFL behavior might be explained by a single QCP at 
negative critical pressure $p_c$ and $B\approx B_+$ or, alternatively 
by an extended quantum critical phase boundary. 
Indeed, high-pressure experiments
indicate that the NFL behavior emanates from a $p$- and $B$-dependent 
line, which is shielded by superconductivity \cite{Ronning_QCP_p}.

% QCP - relation to sc (or the Q phase)
% p-B-T PD
To check whether a quantum critical line is compatible with the
$p$ dependence of $T_c$ and the NFL behavior from our
measurements, we extend the phase diagram to finite pressures
with literature data measured under hydrostatic pressure or on 
Cd-doped CeCoIn$_5$ [Fig.~\ref{fig:PD}(b)]. Although Cd doping
does not generate a noteworthy enlargement of the unit cell,
pressure experiments on CeCoIn$_{5-x}$Cd$_x$ show that a nominal
Cd concentration of 5\% correponds to a negative chemical
pressure of $-0.7\,$GPa \cite{Pham_Cd}. Doping with Cd suppresses
superconductivity and leads to long-range AF order. At $B=0$, a
linear extrapolation of the N\'{e}el temperature $T_\text{N}$ to
positive pressures reveals that, without intervening
superconductivity, AF order would disappear at a pressure where
$T_c$ becomes maximal as indeed shown above for $B=B_c\approx B_+$. 
This is a common feature of many
unconventional superconductors close to AF order and is generally
taken as evidence for superconductivity mediated by magnetic
fluctuations \cite{Monthoux_sc_review,Pleiderer_SC_f}. As shown in
the $T=0$ plane of Fig.~\ref{fig:PD}(b), with increasing $B$, the
$T_c$ maximum follows approximately the origin of NFL behavior to
lower pressures until it reaches at $p=0$ a field that is close
to $B_+$ (or $B_c$). At higher fields, especially at $B^0_{c2}$,
the phase line continues to negative pressures, again in
accordance to our findings. In contrast to the $p$ and $B$
dependence of the NFL behavior, the onset of the discontinuous
superconducting transition and Q phase $B_0(p)$ follows the upper
critical field $B^0_{c2}(p)$ and is still present when, beyond
$p>p_c$, NFL behavior can no longer be observed. Although, at
ambient pressure, $B_0$ and $B_+$ are close to each other, their
qualitatively different pressure dependence gives clear evidence 
that the Q phase can be ruled out as source for the NFL behavior. 
The close relationship between the $T_c$ maximum, the origin of 
NFL behavior, and the endpoint of AF order, on the other hand,
suggest that the quantum critical line corresponds to the AF 
phase boundary. In this case, the available data can be described 
by the tentative phase diagram proposed in Fig.~\ref{fig:PD}(b).

%\section{Summary}
In summary. we performed thermal-expansion measurements at
different magnetic fields to study the pressure and field
dependence of the non-Fermi-liquid behavior in CeCoIn$_5$ which
extends at ambient pressure over a wide field range. Although the
deviations from Fermi-liquid behavior reach a maximum at the
onset of superconductivity at $B^0_{c2}$, the Gr\"uneisen
parameter clearly demonstrates that at $p=0$ no QCP exists up to
a field of 14\,T. Due to the finite distance of the QCP from 
$B^0_{c2}$, the critical behavior changes at low temperatures
toward Fermi-liquid recovery. A combination of our results with
literature data allows to construct a tentative $(p,B,T)$ phase
diagram, in which the NFL behavior originates from a quantum
critical line of the onset of antiferromagnetic order. This line,
emanating from a point of positive pressure on the $B = 0$ axis, 
is hidden by the superconducting dome. It may extend to negative 
pressures at magnetic fields larger than $B^0_{c2}$ as suggested 
by recent experiments on Cd-doped CeCoIn$_5$. 

\begin{acknowledgments}
This work was supported by the Deutsche Forschungsgemeinschaft in 
the form of the Research Unit FOR 960 ``Quantum Phase Transitions''.
Work at Los Alamos was performed under the auspices of the US DOE, 
Office of Basic Energy Sciences, Division of Materials Science 
and Engineering.
\end{acknowledgments}

\end{document}